\documentclass[prl, reprint, superscriptaddress]{revtex4-2}
\pdfoutput=1
\usepackage{braket,dsfont,subfigure,amsfonts,amssymb,bm,graphicx,float,amsmath,hyperref,amsthm,txfonts,orcidlink}
\usepackage{mathrsfs, xcolor}
\usepackage{enumitem}
\newcommand{\Tr}{{\rm Tr}} 
\newcommand{\llvert}{\left|\left|}
\newcommand{\rrvert}{\right|\right|}
\newcommand{\rrangle}{\rangle\!\rangle}
\newcommand{\llangle}{\langle\!\langle}

\newcommand{\llpipe}{|}

\newcommand{\sket}[1]{\ensuremath{\llpipe#1\rrangle}}

\newcommand{\sbraket}[1]{\ensuremath{\llangle#1\rrangle}}
\theoremstyle{definition}

\begin{document}
\title{Locally purified density operators for noisy quantum circuits}
\author{Yuchen Guo~\orcidlink{0000-0002-4901-2737}}
\affiliation{State Key Laboratory of Low Dimensional Quantum Physics and Department of Physics, Tsinghua University, Beijing 100084, China}
\author{Shuo Yang~\orcidlink{0000-0001-9733-8566}}
\email{shuoyang@tsinghua.edu.cn}
\affiliation{State Key Laboratory of Low Dimensional Quantum Physics and Department of Physics, Tsinghua University, Beijing 100084, China}
\affiliation{Frontier Science Center for Quantum Information, Beijing 100084, China}
\affiliation{Hefei National Laboratory, Hefei 230088, China}\begin{abstract}
    Simulating open quantum systems is essential for exploring novel quantum phenomena and evaluating noisy quantum circuits.
    In this Letter, we address the problem of whether mixed states generated from noisy quantum circuits can be efficiently represented by locally purified density operators (LPDOs).
    We map an LPDO of $N$ qubits to a pure state of size $2\times N$ defined on a ladder and introduce a unified method for managing virtual and Kraus bonds.
    We numerically simulate noisy random quantum circuits with depths up to $d=40$ using fidelity and entanglement entropy as accuracy measures.
    LPDO representation proves to be effective in describing mixed states in both quantum and classical regions but encounters significant challenges at the quantum-classical critical point, limiting its applicability to the quantum region exclusively.
    In contrast, the matrix product operator (MPO) successfully characterizes the entanglement trend throughout the simulation, while truncation in MPOs breaks the positivity condition required for a physical density matrix.
    This work advances our understanding of efficient mixed-state representation in open quantum systems and provides insights into the entanglement structure of noisy quantum circuits.
\end{abstract}
\maketitle
\textit{Introduction.\textemdash}
Simulating open quantum systems is crucial for theoretical and practical advancements~\cite{Breuer2007, Rivas2012, Hofer2017, Cattaneo2021, Schlimgen2021, Liu2021, Kamakari2022}, as it enables the investigation of fascinating quantum phenomena in finite-temperature or dissipative systems~\cite{Weiss2012, Kessler2012, Walter2014, Xu2014, KimchiSchwartz2016, Kessler2021, Groot2022, Ma2023A, Ma2023B, Zhang2023}, and plays a crucial role in evaluating the performance of noisy quantum circuits~\cite{Nielsen2010, Bremner2017, Preskill2018, Sarovar2020, Luepke2020, Cheng2021, Cattaneo2023, Torre2023}.
However, dealing with open quantum systems of large size is a formidable challenge due to the exponential growth of the density operator space.
Efficiently representing these mixed states is crucial for the accurate simulation and analysis of noisy quantum circuits.

Traditional tensor network (TN) family~\cite{Verstraete2008, Orus2014, Bridgeman2017, Cirac2021, Bai2022}, which includes matrix product states (MPS)~\cite{Verstraete2006A, PerezGarcia2007, Schollwock2011} and projected entangled pair states (PEPS)~\cite{Verstraete2006B, Schuch2007, Schuch2010, Cirac2011, Schuch2013, Yang2014, Yang2015}, provides intuitive understanding and a compact representation of the entanglement structure in many-body pure state with only a polynomial number of variational parameters and computational costs as the system size grows.
In the realm of simulating open quantum systems, the concept of locally purified density operators (LPDOs)~\cite{Verstraete2004, Zwolak2004, Cuevas2013} has found applications in the study of one-dimensional (1D) open systems governed by master equations~\cite{Werner2016}, simulating noisy quantum circuits~\cite{Cheng2021}, quantum state or process tomography~\cite{Guo2022, Guo2023A, Li2023B, Torlai2023}, and topological quantum matter in open systems~\cite{Guo2024}.
This raises the question of whether a mixed state can be efficiently represented by an LPDO, where the absence of an analytical conclusion hinders the reliability of associated methods and algorithms.

In this Letter, we address this challenge by mapping an LPDO of $N$ qubits to a pure state of size $2\times N$ defined on a ladder, where the implementation of quantum gates and noise channels follows a similar framework.
This unified approach facilitates the simultaneous management of both virtual and inner bonds, leading to the emergence of a critical scaling formula of circuit depth for an efficient LPDO representation, which constitutes the main theoretical contribution of this paper.
To verify this unified framework, we perform numerical simulations involving random noisy quantum circuits with a depth of up to $d=40$.
We evaluate the accuracy of capturing complex dynamics using fidelity and entanglement entropy (EE) as measures.
Throughout the simulations, we observe two well-defined dynamic regions: a quantum region where quantum entanglement continues to accumulate, and a classical region where the system gradually becomes fully depolarized, consistent with previous numerical results~\cite{Noh2020, Li2023A, Zhang2022A, Zhang2022B}.
Both regions allow for an accurate LPDO approximation.
However, the transition between these regions, the quantum-classical crossover point, poses a significant challenge to the classical simulation process, which hinders the simulation of dynamics beyond the critical point.

\textit{LPDO representation for mixed states.\textemdash}
\begin{figure*}
    \includegraphics[width=0.74\linewidth]{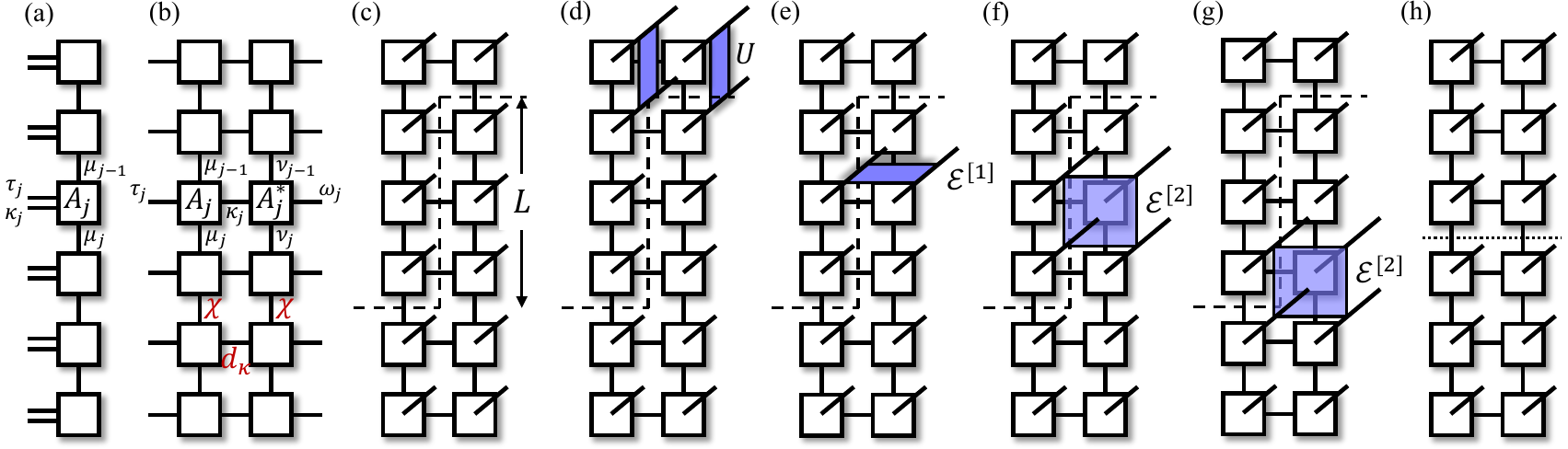}
    \caption{(a) Purified state of an LPDO. (b) Mixed state in its LPDO representation. (c) The corresponding supervector on a ladder. (d-g) Implementation of a quantum circuit in a mixed state. (d) Implementation of a two-qubit gate. (e) Implementation of single-qubit noise. (f-g) Implementation of two-qubit noise. (h) Division for half-cut entanglement entropy used in numerical simulations.}
    \label{Fig: Ladder}
\end{figure*}
We start from an MPS shown in Fig.~\ref{Fig: Ladder}(a) with two physical indices at each site
\begin{align}
    \ket{\psi} = \sum_{\{\bm{\tau}, \bm{\kappa}\}}\sum_{\{\bm{\mu}\}}\prod_{j=1}^N[A_j]^{\tau_j, \kappa_{j}}_{\mu_{j-1}, \mu_{j}}
    \ket{\tau_1,\kappa_1, \cdots,\tau_N, \kappa_N},
\end{align}
where $\bm{\mu}$ denote virtual indices with dimension $\chi$ and $\bm{\kappa}$ represent Kraus indices with dimension $d_{\kappa}$.
Tracing out all the Kraus indices yields the following mixed state.
\begin{align}
    \begin{aligned}
    \rho = \Tr_{\kappa}{\left[\ket{\psi}\hspace{-1mm}\bra{\psi}\right]}
    = &\sum_{\{\bm{\tau}, \bm{\omega}\}}\sum_{\{\bm{\mu}, \bm{\nu}, \bm{\kappa}\}}\prod_{j=1}^N[A_j]^{\tau_j, \kappa_{j}}_{\mu_{j-1}, \mu_{j}}[A_j^*]^{\omega_j, \kappa_{j}}_{\nu_{j-1}, \nu_{j}} \\ &\ket{\tau_1,\cdots,\tau_N}\hspace{-1mm}\bra{\omega_1,\cdots,\omega_N},
    \end{aligned}
\end{align}
as depicted in Fig.~\ref{Fig: Ladder}(b).
Similar to the well-established connection between the locality of interaction and the efficient MPS representation, local purification also requires the locality between system and environment, where system qubits only interact with adjacent ancillae~\cite{SeeSM}.

However, although the virtual and Kraus indices are believed to be related to quantum entanglement and classical mixture, respectively~\cite{Cheng2021}, this interpretation is less straightforward than in the MPS formalism.
In the context of MPS, virtual indices embody the Schmidt decomposition between different subsystems, and an entanglement area law ensures a constant bond dimension $D$~\cite{Verstraete2006A, Bravyi2006, Hastings2007, Eisert2010}.
Moreover, from the purification perspective, we can see that $d_{\kappa}$ equal to the physical bond dimension $d_p$ is sufficient to represent any mixed state exactly, but requires a large $\chi$ that may grow exponentially with the system size.
However, choosing a larger $d_{\kappa}$ can reduce $\chi$ and overall complexity while maintaining accuracy. 
This implies that these two inner indices should be considered together, as illustrated in the following discussion.

A general mixed state allows for an eigenvalue decomposition as $\rho = \sum_k{\lambda_k\ket{\psi_k}\hspace{-1mm}\bra{\psi_k}}$, which naturally corresponds to an unnormalized supervector in double space $\sket{\rho} = \sum_k{\lambda_k\ket{\psi_k}\otimes\ket{\psi_k^*}}$.
In this sense, a one-dimensional (1D) mixed state with $N$ qubits is converted into a quasi-1D system of size $2\times N$.
In particular, a density operator $\rho$ represented by an LPDO
can be viewed as a $2\times N$ state on a ladder, as shown in Fig.~\ref{Fig: Ladder}(c).
Therefore, an efficient LPDO representation for the original mixed state requires the corresponding supervector to satisfy the entanglement area law. 

\textit{Entanglement dynamics in noisy quantum circuits.\textemdash}
Here, we investigate the entanglement dynamics of Kraus and virtual indices in a general noisy quantum circuit. 
As depicted in Fig.~\ref{Fig: Ladder}(c), a division line of length $L$ splits the entire state into two regions, requiring an $O(L)$ scaling of the entanglement entropy between two subsystems to ensure an efficient tensor network representation.
Therefore, it is important to study how these basic components that make up a noisy quantum circuit, namely unitary gates and noise channels, contribute to this bipartite entanglement measure.
It should be noted that to obtain a reasonable estimate of the von Neumann entanglement entropy, the supervector should be renormalized as $\sbraket{\rho|\rho} = \Tr{\left[\rho^2\right]}=1$.

The detailed analysis for the roles of different gates and noise channels is provided in Supplemental Material~\cite{SeeSM}, where the results are summarized as follows.
\begin{itemize}[leftmargin = 3mm, itemsep=-1mm, topsep=0mm]
    \item \textbf{Single-qubit unitary gates.} These gates do not alter the entanglement structure, i.e., $\Delta S = 0$.
    \item \textbf{Two-qubit unitary gates.} One possible position for a two-qubit unitary gate to cross the division line is illustrated in Fig.~\ref{Fig: Ladder}(d), suggesting that a layer of two-qubit gates can increase entanglement by at most $\Delta S \leq 2C_U$.
    \item \textbf{Single-qubit noise.} A single-qubit noise channel (Fig.~\ref{Fig: Ladder}(e)) exhibits little difference from the two-qubit unitary gate within this framework.
    Consequently, a layer of single-qubit noise channels induces a change in entanglement that satisfies $\Delta S\leq LC_{\mathcal{E}}^{[1]}$.
    \item \textbf{Two-qubit noise.} Similarly, a layer of two-qubit noise channels will intersect the division by $L+1$ times, as shown in Fig.~\ref{Fig: Ladder}(f-g), resulting in an entanglement increase of $\Delta S\leq (L+1) C_{\mathcal{E}}^{[2]}$.
\end{itemize}
Here, $C_U$ represents a constant of order $O(1)$ describing the capacity of a unitary gate to generate entanglement in a normal quantum state.
Furthermore, $C_{\mathcal{E}}^{[i]}$ is proportional to the error rate $\varepsilon^{[i]}$ of the corresponding $i$-qubit gates.

In summary, for a quantum circuit with a staggered arrangement of single-qubit and two-qubit gates (where the two-qubit gates form a brick-wall structure)~\cite{SeeSM, Guo2022}, the entanglement growth under circuit depth $d$ can be estimated as
\begin{align}
    \Delta S\leq \frac{d}{2}(2C_U + LC_{\mathcal{E}}^{[1]} + (L+1) C_{\mathcal{E}}^{[2]}) \sim O(d\varepsilon L),\label{Equ: area_law}
\end{align}
which is satisfied for large $L$.
Consequently, the supervector $\sket{\rho}$ for a mixed state generated from a noisy circuit with a depth scaling lower than $d\sim \varepsilon^{-1}$ satisfies $S\lesssim O(L)$.
This depth scaling ensures an efficient tensor network representation for $\sket{\rho}$, and thus an efficient LPDO representation for $\rho$.
In other words, we expect a failure of the LPDO simulation when the circuit is deeper than the scaling behavior $d\sim \varepsilon^{-1}$ that is irrelevant to the concrete unitary gates used in the circuit.
Moreover, this theoretical prediction applies to a general gate configuration if considering the relative density of gates $p$ compared to the brick-wall case to define an effective error rate $p\varepsilon$, as demonstrated in the following numerical experiments.

\textit{Numerical simulations for noisy quantum circuits.\textemdash}
\begin{figure*}
    \includegraphics[width=\linewidth]{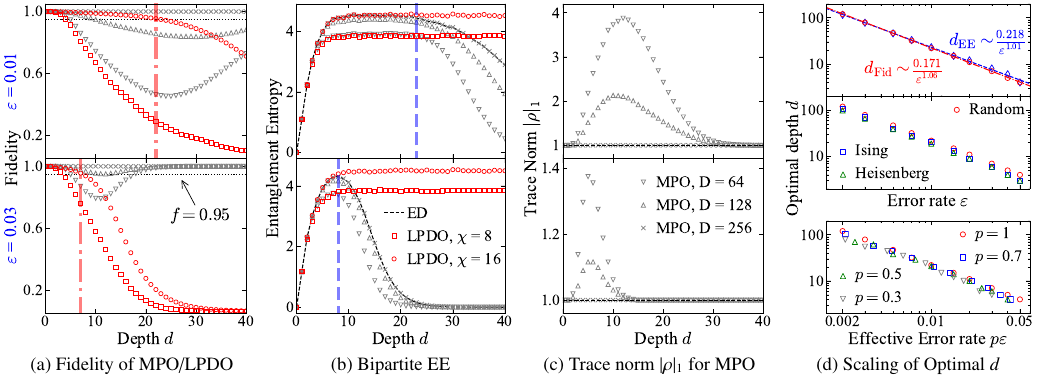}
    \caption{(a-c) Dynamics of noisy quantum circuits with $N=8$ for ED, MPO with different $D$, and LPDO with $d_{\kappa}=2$ and different $\chi$, sharing the same legend.
    The upper and lower panels show the results for $\varepsilon=0.01$ and $\varepsilon=0.03$, respectively.
    (a) Fidelity between ED and MPO/LPDO results. The red dashed lines indicate the positions where $f_{\rm LPDO}$ ($\chi=16$, $d_{\kappa}=2$) drops below $0.95$.
    (b) Bipartite entanglement entropy for different methods.
    The blue dashed lines indicate the positions where $\rm EE_{ED}$ and $\rm EE_{LPDO}$ ($\chi=16$, $d_{\kappa}=2$) deviate by more than $3\%$.
    (c) Trace norm $|\rho|_1$ for MPO with different $D$.
    (d) The `optimal depth' for LPDO simulation under different $\varepsilon$.
    The upper panel: $d_{\rm opt}$ determined by fidelity and EE for random gates.
    The middle panel: $d_{\rm opt}$ determined by the fidelity for different gates, including random gates and time evolution of two spin models.
    The lower panel: $d_{\rm opt}$ determined by the fidelity for random gates, where each gate is randomly preserved with probability $p$.}
    \label{Fig: EE}
\end{figure*}
In our numerical simulations, we use brick-wall circuits with Haar-random two-qubit gates~\cite{Nahum2017, Fisher2023, SeeSM} accompanied by two-qubit depolarizing noise.
Initially, we focus on systems with $N=8$ qubits whose entire dynamics can be exactly simulated as a benchmark.
We evaluate supervector fidelity, defined as $f\left(\sket{\rho_1}, \sket{\rho_2}\right) \equiv \sbraket{\rho_1|\rho_2}$ with $\sbraket{\rho_i|\rho_i} = 1$~\cite{SeeSM}, between matrix product operator (MPO) or LPDO and exact diagonalization (ED) in Fig.~\ref{Fig: EE}(a) for different two-qubit error rates $\varepsilon$.
The truncation of MPO follows the conventional method~\cite{Noh2020}, while we improve the LPDO truncation method to enhance accuracy and robustness~\cite{SeeSM}.

First, for a typical $\varepsilon=0.01$ of state-of-the-art quantum hardware, we observe a region where LPDO with $\chi=16$ and $d_{\kappa}=2$ (red circles) exhibits expressive power comparable to that of an MPO with $D = \chi^2=256$ (grey crosses, exact for $N=8$) and better than $D=\chi^2/2=128$ (grey triangles) up to depth $d=33$, while the LPDO structure consumes significantly fewer computational resources ($d_p^2D^2$) compared to MPO ($d_pd_{\kappa}\chi^2$).
Meanwhile, this region is smaller for stronger noise when comparing the upper and lower panels of Fig.~\ref{Fig: EE}(a).
Therefore, our results underscore the importance of selecting an appropriate ansatz when simulating noisy circuits at different error rates $\varepsilon$.

Nevertheless, it is crucial to note that with the accumulation of noise, an LPDO can no longer accurately capture the exact trajectories due to large decoherence effects after a characteristic depth.
This `optimal depth', defined as the point where the fidelity between ED and LPDO with $\chi=16$ and $d_{\kappa}=2$ drops below $f=0.95$, is marked by red dashed lines in Fig.~\ref{Fig: EE}(a) and fitted as $d_{\rm Fid}\sim 0.171/\varepsilon^{1.06}$ in the upper panel of Fig.~\ref{Fig: EE}(d).
These results reveal the failure of the LPDO simulation beyond this optimal depth, validating our theoretical analysis and prediction based on Eq.~\eqref{Equ: area_law}.
In addition, this result is independent of the concrete two-qubit unitary gates and their configuration, as demonstrated by the collapse across different gates, including random gates and time evolution of two spin models, as well as different configurations achieved by randomly preserving each gate with probability $p$ (resulting in an effective error rate of $p\varepsilon$~\cite{SeeSM}), as shown in the middle and lower panels of Fig.~\ref{Fig: EE}(d), respectively.
This suggests the universality of our theoretical formalism in Eq.~\eqref{Equ: area_law},  indicating that the optimal depth is determined solely by the effective noise strength, regardless of the unitary gates involved.

\begin{figure*}
    \includegraphics[width=\linewidth]{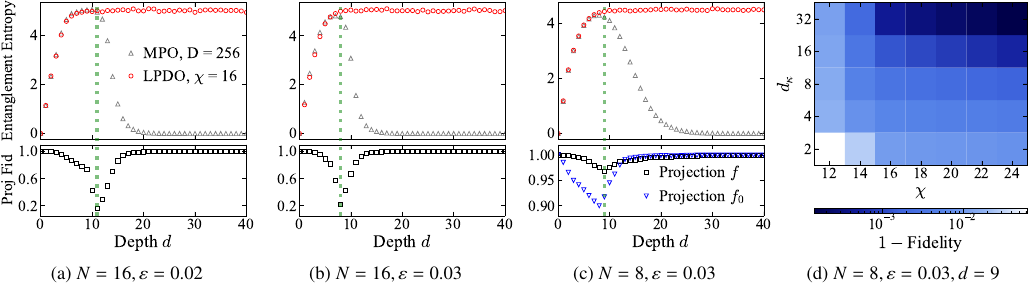}
    \caption{Projection fidelity for projection from an MPO to an LPDO compared with entanglement entropy dynamics simulated by both structures.
    (a-c) The upper panels show the EE for noisy circuits with different $N$ and different $\varepsilon$.
    The lower panels show the fidelity of projecting an MPO with $D=256$ onto an LPDO with $\chi=16$ and $d_{\kappa}=2$.
    The green dotted line indicates the position of minimum fidelity.
    The conventional fidelity $f_0$ for mixed states and the supervector fidelity $f$ are compared in (c).
    (d) Projection from an MPO with $D=256$ generated from noisy circuits with $N=8$, ${\rm depth}=9$, and $\varepsilon=0.03$ to an LPDO with different $\chi$ and $d_{\kappa}$.
    }
    \label{Fig: Proj}
\end{figure*}

To understand the physical interpretation for this optimal depth, we compare the bipartite EE of the supervector $\sket{\rho}$ (divided into half of the chain, as shown in Fig.~\ref{Fig: Ladder}(h)) for the simulations of ED, MPO, and LPDO in Fig.~\ref{Fig: EE}(b).
We observe an increasing-decreasing trend of EE from ED simulation, dividing the entire dynamics into two regions:
(1) a quantum region where quantum entanglement gradually accumulates, potentially offering quantum advantages for systems with much larger sizes; 
(2) a classical region dominated by noise effects, leading to highly depolarized and mixed systems, consistent with the numerical results in recent works~\cite{Noh2020, Zhang2022A, Li2023A}.
MPO simulation, even after truncation, can qualitatively capture this trend, while LPDO only accurately describes EE dynamics in the quantum region.
In other words, the failure of LPDO occurs just at the critical point when the circuit EE starts to decrease from its maximal value and transitions into the classical region.
We mark this critical point, defined as the point where EE of LPDO with $\chi=16$ and $d_{\kappa}=2$ and ED differ by more than $3\%$, by blue dashed lines in Fig.~\ref{Fig: EE}(b) and fitted in the upper panel of Fig.~\ref{Fig: EE}(d), consistent with the optimal point defined by the fidelity shown before.
This implies that the intricate interplay between entanglement and noise in the quantum-classical crossover region leads to the failure of the LPDO simulation.

On the other hand, while MPO simulation effectively captures the overall trend of EE dynamics in both regimes, it may violate the positivity condition for a physical density matrix that cannot be imposed on the local tensor of an MPO.
To provide a comprehensive assessment of MPO performance, we calculate the trace norm $|\rho|_1$ for normalized MPO with $\mathrm{Tr}(\rho)=1$.
This measure, which is the summation of all singular values, will exceed $1$ if there are negative components in the density operator.
The results presented in Fig.~\ref{Fig: EE}(c) reveal that after truncation, which is inevitable for simulating larger systems with finite depth $D$, the MPO fails to preserve the positivity condition.
Furthermore, we observe that the maximal `negativity' occurs prior to the critical point, suggesting a potential causal relationship between these two issues and offering more insight into the difficulty of LPDO at the critical point.

In our previous analysis, we primarily focused on the entanglement structure when deriving the critical scaling, which serves as a necessary condition for LPDO representation.
However, due to its inherent structure that aims to preserve local positivity, the efficacy of the LPDO representation even with infinite $d_{\kappa}$ may be limited compared to MPO that lacks such constraints.
Therefore, we anticipate that failure of the LPDO structure right follows the peaks of MPO negativity, where there is a crucial trade-off between the accuracy of observables and the positivity of output states.
This conflict cannot be alleviated only by increasing $d_{\kappa}$, as illustrated in Fig.~S4~\cite{SeeSM}.
Fortunately, with the development of quantum hardware, only dynamics in the quantum region can demonstrate a quantum advantage, where LPDO is capable of capturing long-time dynamics of interest and far exceeds MPO in terms of both accuracy and efficiency, as demonstrated in the `weak-noise' case (Fig.~S6~\cite{SeeSM}).

\textit{Projection to an LPDO.\textemdash}
To further explore the properties of these dynamical critical points and explain the limitations of the LPDO simulation in this context, we propose a gradient descent algorithm to project an MPO onto an LPDO.
To be more specific, we aim to optimize the following loss function
\begin{align}
    \Theta = \llvert\rho-\rho^{\prime}\rrvert_F^2 = \Tr{\left[\rho^2 + \rho^{\prime2} -2\rho\rho^{\prime}\right]},
\end{align}
where $\rho$ represents the original MPO and $\rho\prime$ denotes the approximated LPDO.
To obtain an approximated LPDO $\rho^{\prime}$ with minimal loss $\Theta$, we perform a gradient descent using Adam optimizer~\cite{Kingma2017}.

Fig.~\ref{Fig: Proj}(a-c) plot the projection fidelity from the simulated MPO at each depth with $D=256$ to an LPDO with fixed $\chi = 16$ and $d_{\kappa}=2$ for different $N$ and $\varepsilon$ (distinct from the fidelity for LPDO simulation shown in Fig.~\ref{Fig: EE}(a)), along with the directly simulated LPDO EE with the same $d_{\kappa}$ and $\chi$ for visualization of the critical point discussed previously.
Here, we consider both the density matrix fidelity $f_0$ and the supervector fidelity $f$ Fig.~\ref{Fig: Proj}(c) for $N=8$~\cite{SeeSM}, where the results clearly reveal a similar trend for these two fidelity measures.
Minimum fidelity is observed around the critical point where the EE estimated by MPO and LPDO deviate, indicating a fundamental limitation of LPDO representation in that region.
Furthermore, the deviation from the correct trajectory for the dynamics near the critical point hinders LPDO simulation at and after that point because the LPDO trajectory cannot be directly corrected back to its original dynamics once destroyed.
Therefore, although the mixed states in the classical region can be well approximated by an LPDO, the LPDO simulation cannot reach those states, leading to the difference between Fig.~\ref{Fig: EE}(a) and Fig.~\ref{Fig: Proj}.

Meanwhile, the overall trend of projection fidelity is quite similar when comparing Fig.~\ref{Fig: Proj}(b) and \ref{Fig: Proj}(c), but the values around the critical point for $N=16$ are much lower than $N=8$, where an exact MPO representation is available.
This unsatisfactory fidelity is attributed to the truncation of MPO during the simulation that introduces non-positive parts as demonstrated in Fig.~2(c).
This indicates a difference between LPDO with $\chi$ and MPO with $D=\chi^2$ in their representative capacity for highly compressed mixed states, where adherence to the positivity condition is most challenging.
This highlights the difficulty for LPDO in accurately capturing the dynamics under a stringent positivity constraint.

Finally, we compare the expressive capacity of LPDO with different $\chi$ and $d_{\kappa}$ for noisy quantum states.
To avoid the truncation for the MPO representation that may lead to the nonphysical properties mentioned above, we adopt $N=8$ and directly implement the projection from the exact density matrices generated from the noisy circuits with $\varepsilon=0.03$ and $d = 9$ to LPDO, just at the critical point between the quantum and classical regions.
From the projection fidelity shown in Fig.~\ref{Fig: Proj}(d), we observe that, for instance, an LPDO with $\chi=24$ and only $d_{\kappa}=4$ can possess a representative power similar to that of an LPDO with $\chi=14$ but much larger $d_{\kappa}$, at least for these quantum states generated from typical noisy quantum circuits of interest.
This implies that the difference between virtual indices and Kraus indices is not absolute, and a unified scheme to treat them according to our proposal is reasonable.

\textit{Conclusion and discussion.\textemdash}
In conclusion, our investigation into the simulation of open quantum systems through the LPDO representation has clarified its capacity and limitations when dealing with the dynamics of noisy quantum circuits.
With theoretical predictions and numerical simulations, we observe a critical point between the quantum and classical regions, which remains a challenge for LPDO representation, as evidenced by the divergence in EE from exact results and a minimal fidelity for the projection.

Several implications of our study emerge.
Firstly, the universality of the critical scaling embodied in Eq.~\eqref{Equ: area_law}, underscored by the consistency across various unitary gates and circuit configurations, offers a strategic approach to estimate the optimal depth for any quantum circuit.
Specifically, the corresponding scaling coefficient can be determined based on an arbitrary error rate $\varepsilon$, by, e.g. conducting an MPO simulation one time, and then the quantum region conducive to the LPDO simulation can be identified.
As a consequence, continued MPO simulation becomes unnecessary, as LPDO suffices to characterize the dynamics in the quantum region, while the classical regime diminishes in relevance with little quantum advantage.
Meanwhile, our results delineate the application scope of these LPDO-based quantum state and process tomography methods~\cite{Guo2023A, Li2023B, Torlai2023} from the optimal depth predicted.

Further exploration and characterization of the interplay between noise effects and quantum entanglement, especially in the crossover between quantum and classical regions, are essential~\cite{Guo2023C}.
The possible connections to these measurement or noise-induced phase transitions~\cite{Li2018, Skinner2019, Li2019, Vasseur2019, Yang2022, Guo2023B} or other dynamical phase transitions~\cite{Diehl2010, Zhang2017, Heyl2018, Muniz2020, Marino2022} are also interesting.
In summary, the challenges and opportunities identified in this study provide a foundation for further research into the efficient representation and simulation of open quantum states, offering potential advances in quantum information processing and quantum computing.

\begin{acknowledgments}
    This work is supported by the National Natural Science Foundation of China (NSFC) (Grant No. 12174214, No. 12475022, and No. 92065205) and the Innovation Program for Quantum Science and Technology (Grant No. 2021ZD0302100).
\end{acknowledgments}

\bibliography{ref}

\appendix
\onecolumngrid
\renewcommand{\thesection}{S-\arabic{section}} \renewcommand{\theequation}{S%
\arabic{equation}} \setcounter{equation}{0} \renewcommand{\thefigure}{S%
\arabic{figure}} \setcounter{figure}{0}
\newpage
\section*{Supplemental Material}

In this supplemental material, we provide more details on the analysis of entanglement dynamics under noisy quantum circuits, the conversion between MPO and LPDO, the truncation method of LPDO, the fidelity of supervectors, the circuit configuration in numerical simulations, and additional numerical results.

\section{Entanglement dynamics under noisy quantum circuits}
In this section, we conduct a comprehensive investigation of the impacts of each component in a typical noisy quantum circuit on the entanglement structure of an LPDO.
It is obvious that a single-qubit gate will not affect any entanglement structure, so we proceed directly to the analysis of two-qubit gates.

\textbf{1. Two-qubit unitary gates.} Consider a conventional two-qubit gate applied to the mixed state as $U\rho U^{\dagger}$, where the corresponding transformation in the supervector is expressed as
\begin{align}
    U\otimes \overline{U}\sket{\rho}.
\end{align}
The only scenario for a two-qubit gate to change the entanglement of $\sket{\rho}$ is when the unitary crosses the division line, as illustrated in Fig.~1(d).
Suppose that the potential capacity for the unitary to generate entanglement in a normal quantum state is denoted as $C_{U}$, (e.g., $C_{U}$ for a CNOT gate is $C_{U} = \ln{2}$), then it can be directly shown that the entanglement increase of the supervector also follows the conventional rule.
Consequently, if a layer of two-qubit gates is applied to the quantum state, with one gate acting on each nearest-neighbor pair, the entanglement entropy of the supervector across the division line shown in Fig.~1(d) satisfies that
\begin{align}
    \Delta S \leq 2C_U.
\end{align}

\textbf{2. Single-qubit noise.} General quantum noise can be expanded in its operator-sum representation as follows
\begin{align}
    \mathcal{E}^{[1]}\left(\rho\right) = \sum_k E_k\rho E_k^{\dagger}.
\end{align}
As illustrated in the following example, such a single-qubit noise channel (Fig.~1(e)) exhibits little difference from the two-qubit unitary gate discussed before from the supervector perspective.
Here we consider a single-qubit Pauli error as an example, $\mathcal{E}\left(\rho\right) = (1-\varepsilon)\rho + \varepsilon \sigma^x\rho\sigma^x$, whose effect on the supervector is given by
\begin{align}
    \mathcal{E}\sket{\rho} \propto (1-\varepsilon)\sket{\rho} + \varepsilon\sigma^x\otimes \overline{\sigma^{x}}\sket{\rho}.
\end{align}
This behaves just like an entangled gate (non-unitary though) applied across two subsystems, a fact that can be also inferred from the similarity between Fig.~1(d) and (e).
As a result, if each qubit undergoes a single-qubit noise, we expect that the change in entanglement entropy satisfies that
\begin{align}
    \Delta S\leq LC_{\mathcal{E}}^{[1]},
\end{align}
where $C_{\mathcal{E}}^{[1]}\sim a\varepsilon$ with $a$ being a constant of the order $O(1)$.

\textbf{3. Two-qubit noise.} Finally, we examine the effect of a two-qubit noise channel on supervector entanglement.
We again consider a two-qubit Pauli noise as an illustrative example, i.e., $\mathcal{E}^{[2]}(\rho) = (1-\varepsilon)\rho + \varepsilon\left(\sigma^x_i\otimes\sigma^x_j\right)\rho\left(\sigma^x_i\otimes\sigma^x_j\right)$ for adjacent sites $\braket{i, j}$.
The corresponding transformation of the supervector is
\begin{align}
    \mathcal{E}\sket{\rho} \propto (1-\varepsilon)\sket{\rho} + \varepsilon\sigma^x_i\otimes\sigma^x_j\otimes\overline{\sigma^x_i}\otimes\overline{\sigma^x_j}\sket{\rho},
\end{align}
where two cases arise as shown in Fig.~1(f) and (g).
In Fig.~1 (f), the two qubits that support the noise channel lie within the divided subsystem, while the pair $\braket{i, j}$ crosses the division in Fig.~1(g).
Nevertheless, it can be readily verified that these two cases will lead to the same increase in the entanglement entropy of $\sket{\rho}$ with $\Delta S\leq C_{\mathcal{E}}^{[2]}$.
This means that a layer of two-qubit noise, where a total of $L+1$ local channels cross the division, will induce an entanglement increase of
\begin{align}
    \Delta S\leq (L+1) C_{\mathcal{E}}^{[2]}.
\end{align}
Similarly, $C_{\mathcal{E}}^{[2]}\sim b\varepsilon$ with $b$ being a constant of order $O(1)$.

\section{The conversion between MPO and LPDO}
\begin{figure}[H]
    \centering
    \includegraphics[width=0.65\linewidth]{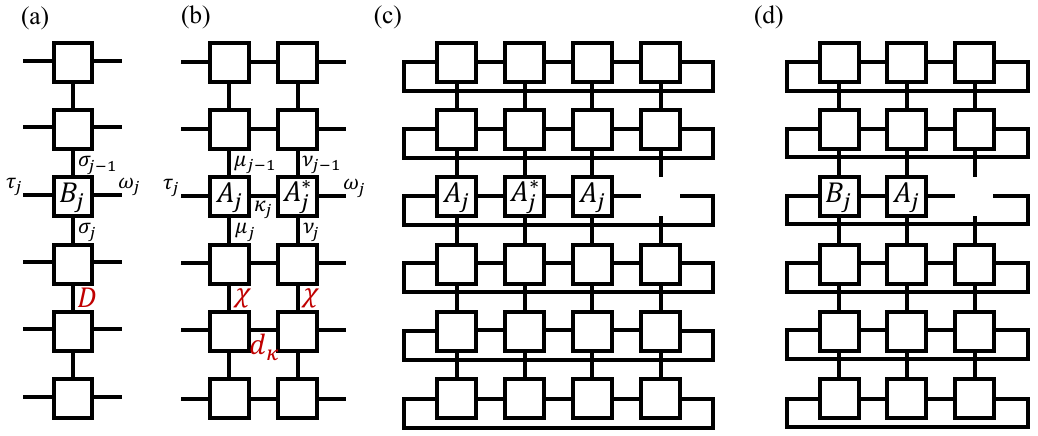}
    \caption{MPO and LPDO representation for mixed states.
    (a) A mixed state represented by an MPO.
    (b) A mixed state represented by an LPDO.
    (c) The environment for $\Tr{\left[\rho^{\prime}\left(\partial\rho^{\prime}/\partial A_j^*\right)\right]}$.
    (d) The environment for $\Tr{\left[\rho\left(\partial\rho^{\prime}/\partial A_j^*\right)\right]}$.}
    \label{Fig: MPO}
\end{figure}

The LPDO structure in Fig.~\ref{Fig: MPO}(b) is proposed in Ref.~\cite{Verstraete2004} as a natural generation of MPS to represent mixed states, which is sometimes also known as the locally purified form of the matrix product density operator (MPDO) or simply MPDO.
An LPDO density matrix is written as
\begin{align}
    \rho = \sum_{\{\bm{\tau}, \bm{\omega}\}}\sum_{\{\bm{\mu}, \bm{\nu}, \bm{\kappa}\}}\prod_{j=1}^N[A_j]^{\tau_j, \kappa_{j}}_{\mu_{j-1}, \mu_{j}}[A_j^*]^{\omega_j, \kappa_{j}}_{\nu_{j-1}, \nu_{j}}
    \ket{\tau_1,\cdots,\tau_N}\hspace{-1mm}\bra{\omega_1,\cdots,\omega_N},
\end{align}
where $\bm{\kappa}$ are Kraus indices that represent the environment (or ancillae) to be traced out.
In this sense, $\ket{\psi}$ serves to purify $\rho$, where an ancillary degree of freedom (the Kraus index) is attached to the physical index via a local tensor at each site.
Therefore, LPDO provides a locally purified form for the density matrix, from which it derives its name.
Notably, an LPDO is guaranteed to be Hermitian and semidefinite positive by design due to its quadratic form with respect to local tensors.

Another commonly adopted TN member to represent a mixed state is the MPO shown in Fig.~\ref{Fig: MPO}(a)
\begin{align}
    \rho = \sum_{\{\bm{\tau}, \bm{\omega}\}}\sum_{\{\bm{\sigma}\}}\prod_{j=1}^N[B_j]^{\tau_j, \omega_{j}}_{\sigma_{j-1}, \sigma_{j}}
    \ket{\tau_1,\cdots,\tau_N}\hspace{-1mm}\bra{\omega_1,\cdots,\omega_N},
\end{align}
which can be obtained by directly contracting the two local tensors at each site, i.e.,
\begin{align}
    [B_j]^{\tau_j, \omega_{j}}_{\sigma_{j-1}, \sigma_{j}} = \sum_{\kappa_j}[A_j]^{\tau_j, \kappa_{j}}_{\mu_{j-1}, \mu_{j}}[A_j^*]^{\omega_j, \kappa_{j}}_{\nu_{j-1}, \nu_{j}}
\end{align}
with $\sigma_j = \left\{\mu_j, \nu_j\right\}$.
In other words, the bond dimension $D$ of an MPO is typically much larger than $\chi$ of an LPDO in the weak-noise region,
In the pure-state limit, an MPS (which corresponds to an LPDO with $d_{\kappa}=1$) with $\chi$ corresponds to an MPO with $D = \chi^2$.
In addition, it is hard to directly impose the constraints of Hermicity and positivity on the construction of an MPO, and to our knowledge, there has been no approach to recover a legible density matrix from a general MPO so far.
Therefore, more and more efforts have been put into the LPDO representation and simulation for open quantum systems.

We have introduced an approach to project any MPO to an LPDO with given $\chi$ and $d_{\kappa}$, or more precisely, to find the LPDO with the smallest distance from the target MPO defined by the Frobenius norm in Eq.~(4) of the main text.
The gradient of each tensor is analytically derived as
\begin{align}
    \frac{\partial \Theta}{\partial A_j^*} = 2\Tr{\left[\left(\rho^{\prime}-\rho\right)\frac{\partial\rho^{\prime}}{{\partial A_j^*}}\right]},
\end{align}
which is shown in Fig.~\ref{Fig: MPO}(c-d) and can be efficiently computed in $O(N)$ time by caching the tensor environment~\cite{Crosswhite2008, Orus2014, SeeSM}.
In each iteration step, we update the local tensor $A_j$ according to the following rule
\begin{align}
    A_{j} \rightarrow A_{j} - \eta \frac{\partial \Theta}{\partial A_{j}^{*}},
\end{align}
where $\eta$ is the learning rate, automatically adjusted using the Adam optimizer~\cite{Kingma2017}.
The hyperparameters in the Adam optimizer are set as $\xi_1 = \xi_2 = 0.8$ and $\epsilon = 10^{-8}$ throughout the numerical simulations in this work.
Importantly, this method also applies to the truncation of an MPO or LPDO, which can be viewed as a projection from an MPO/LPDO with a larger bond dimension to another one with a smaller bond dimension.

\section{The truncation method of LPDO}
The LPDO truncation in previous work~\cite{Werner2016, Cheng2021} is done by directly applying the singular value decomposition (SVD) to local tensors and truncating the singular values for the Kraus indices and virtual indices in sequence.
For instance, all Kraus indices were first truncated via local SVD without considering the tensor environment, and then the virtual indices were truncated site by site in the canonical form in Ref.~\cite{Cheng2021}.
Such a sequence for different indices and the neglect of the environment when truncating $d_{\kappa}$ limit accuracy and robustness.
Here, we introduce a modified three-step compression scheme, where a site-by-site QR and LQ decomposition is first performed to obtain the gauge transformations $L_i$ and $R_i$ on each virtual index.
Next, the projectors for Kraus indices and virtual indices are calculated, respectively, with the standard SVD compression.
Finally, all projectors are applied simultaneously to complete the truncation of LPDO.
\begin{figure}[H]
    \centering
    \includegraphics[width=0.7\linewidth]{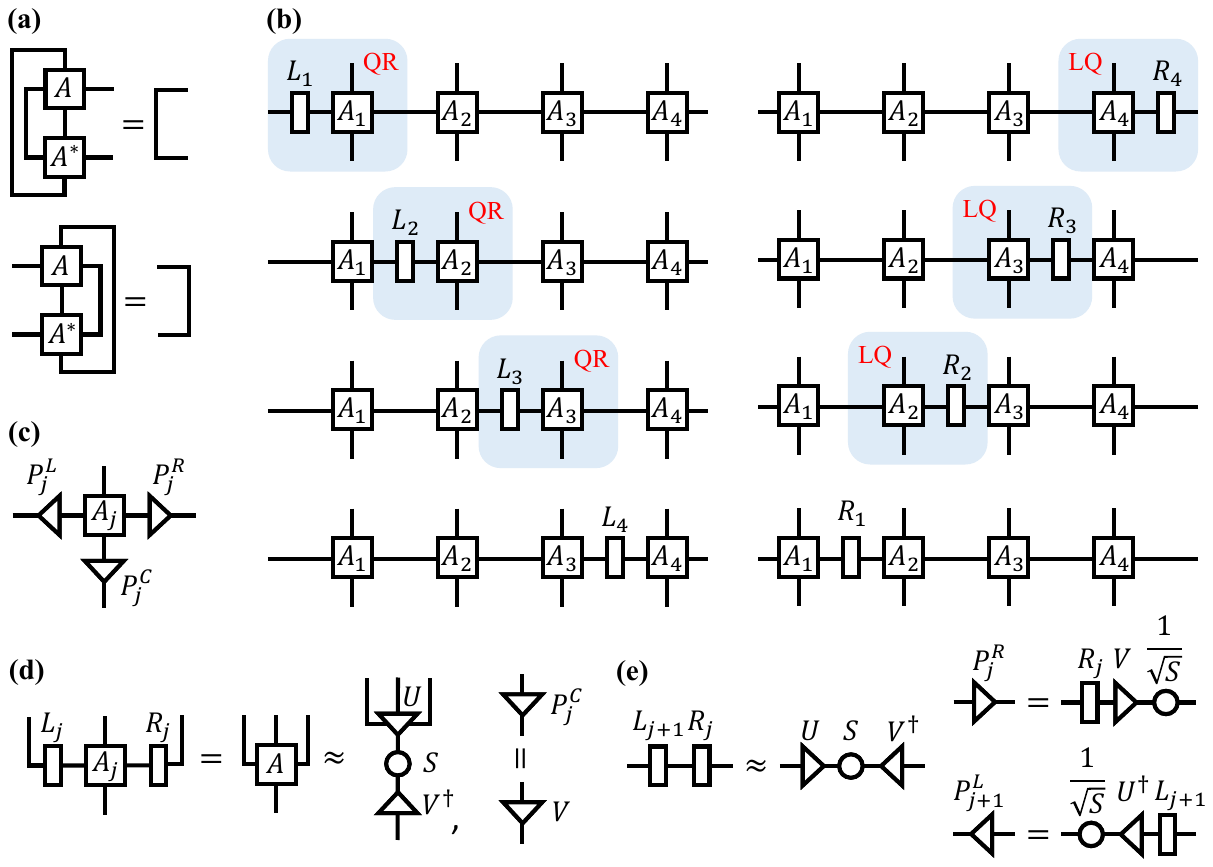}
    \caption{Truncation of LPDO.
    (a) Left and right canonical conditions.
    (b) Site-by-site QR and LQ decomposition.
    (c) Projectors for truncation.
    (d) Construction of Kraus-index projector.
    (e) Construction of virtual-index projector.}
    \label{Fig: Trun}
\end{figure}

The canonical form of an LPDO is naturally generalized from that of an MPS, where both physical indices and Kraus indices are traced when calculating the environment, as shown in Fig.~\ref{Fig: Trun}(a).
First, we implement a left-to-right QR decomposition in Fig.~\ref{Fig: Trun}(b) to obtain the gauge matrices $L_i$ for the left-canonical form, where $L_1 = [1]$, a $1\times 1$ identity matrix.
Similarly, a left-to-right LQ decomposition starts from $R_{N}=[1]$ is performed to calculate $R_i$ for the right-canonical form.
All $L_i$ and $R_i$ are stored for later construction of projectors.
Next, we construct the projectors for both virtual indices $P_j^L, P_j^R$, and Kraus index $P_j^C$ for each local tensor $A_j$ in Fig.~\ref{Fig: Trun}(c) inspired by Ref.~\cite{Yang2017}.
For the Kraus index, we absorb the gauge in virtual indices $L_j$ and $R_j$ into $A_j$ leading to $\tilde{A}_j$ and perform SVD along the vertical direction $\tilde{A}_j\approx USV^{\dagger}$, as shown in Fig.~\ref{Fig: Trun}(d).
The local projector is constructed as $P_j^C = V$.
It can be easily verified that
\begin{align}
    \tilde{A}_jP_j^CP^{C\dagger}_j\tilde{A}_j^{\dagger} \approx USV^{\dagger}VV^{\dagger}VSU^{\dagger} = US^2U^{\dagger},
\end{align}
which does realize the compression of the Kraus index.
Regarding the virtual index, we contract the gauge from two sides and implement SVD, i.e., $L_{j+1}R_j\approx USV^{\dagger}$, and construct the corresponding projectors in Fig.~\ref{Fig: Trun}(e), namely $P_j^R = R_jV\frac{1}{\sqrt{S}}$ and $P_{j+1}^L=\frac{1}{\sqrt{S}}U^{\dagger}L_{j+1}$.
Such projectors satisfy
\begin{align}
    P_j^RP_{j+1}^L = R_jV\frac{1}{\sqrt{S}}\frac{1}{\sqrt{S}}U^{\dagger}L_{j+1}\approx R_j\left(L_{j+1}R_j\right)^{-1}L_{j+1} = I.
\end{align}
We must emphasize that in this process, the Kraus index has not been truncated yet in order to maintain the most accurate tensor environment.
Finally, all projectors $P_j^C$, $P_j^L$, and $P_j^L$ constructed before are simultaneously applied to the local tensor of each site, resulting in a truncated LPDO with smaller $\chi$ and $d_{\kappa}$.

\section{Fidelity between two supervectors}
The fidelity between two normalized pure states is defined as
\begin{equation}
    f\left(\ket{\psi}, \ket{\phi}\right) = \left|\braket{\psi|\phi}\right|^2.
    \label{equ: fidelity}
\end{equation}
This is usually generalized for two mixed states as
\begin{equation}
    f\left({\rho}_1, {\rho}_2\right) = \left(\Tr{\sqrt{\sqrt{{\rho}_1}{\rho}_2\sqrt{{\rho}_1}}}\right)^2
\end{equation}
with normalized $\Tr{\left[{\rho}_1\right]}=\Tr{\left[{\rho}_2\right]}=1$.
However, this definition cannot be directly estimated for two mixed states in their tensor network form, and thus only applies to ED and is intractable for large systems.
In addition, it is even not well defined for density operators violating the positivity condition, such as those truncated MPO.
As an alternative, we consider the fidelity between two supervectors
\begin{equation}
    f\left(\sket{\rho_1}, \sket{\rho_2}\right) \equiv \frac{\sbraket{\rho_1|\rho_2}}{\sqrt{\sbraket{\rho_1|\rho_1} \sbraket{\rho_2|\rho_2}}}
    = \frac{\Tr{\left({\rho}_1{\rho}_2\right)}}{\sqrt{\Tr{\left({\rho}_1^2\right)}\Tr{\left({\rho}_2^2\right)}}},
\end{equation}
which also corresponds to the inner product in the operator space and can be efficiently calculated for both MPOs and LPDOs.
In particular, this alternative definition of fidelity reduces to Eq.~\eqref{equ: fidelity} for pure states.

\section{Circuit configuration in numerical simulations}
\begin{figure}[H]
    \centering
    \includegraphics[width=0.3\linewidth]{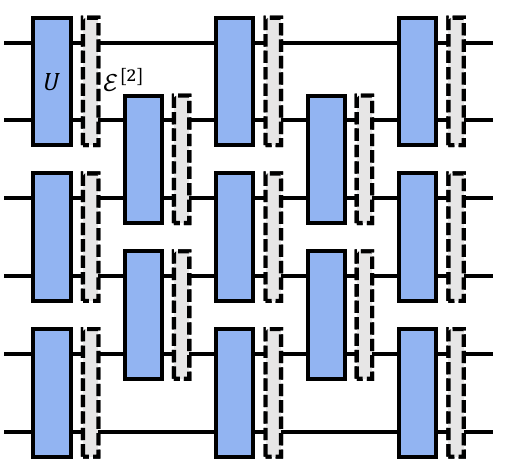}
    \caption{The circuit configuration. Two-qubit gates $U$ follow the random Haar measure, followed by two-qubit depolarizing noise $\mathcal{E}^{[2]}$.}
    \label{Fig: Circuit}
\end{figure}

In the numerical simulations, we use the test circuit shown in Fig.~\ref{Fig: Circuit}, as commonly adopted in various quantum computing tasks~\cite{Temme2017, Noh2020}.
In this circuit, each layer is a tensor product of two-qubit Haar-random gates with a staggered arrangement between adjacent layers.
To be more specific, each local gate $U$ is drawn randomly and independently of all others from the uniform distribution on the unitary group $U(4\times4)$~\cite{Fisher2023}.
We introduce two-qubit depolarizing noise channels after each gate, defined as
\begin{align}
    \begin{aligned}
        \mathcal{E}^{[2]}\left(\rho\right) = \left(1-\frac{16}{15}\varepsilon\right)\rho + \frac{1}{15}\varepsilon\sum_{i,j=0}^{3}\left(\sigma_i\otimes\sigma_j\right)\rho\left(\sigma_i\otimes\sigma_j\right),
    \end{aligned}
\end{align}
where $\varepsilon$ is the two-qubit error rate, typically at the order of $10^{-2}$ for state-of-the-art quantum hardware.

In the lower panel of Fig.~2(d), we randomly discard certain gates, along with their accompanying noise, and retain each gate with a probability $p$. 
This investigation allows us to explore the impact of the gate configuration on our theoretical scaling form.
Here, $p\varepsilon$ serves as an effective error rate, representing the 'density' of noise within a general quantum circuit.
Remarkably, we find that this quantity plays a pivotal role in determining the performance of the LPDO simulation.

\section{Additional numerical results}
\begin{figure}[H]
    \centering
    \includegraphics{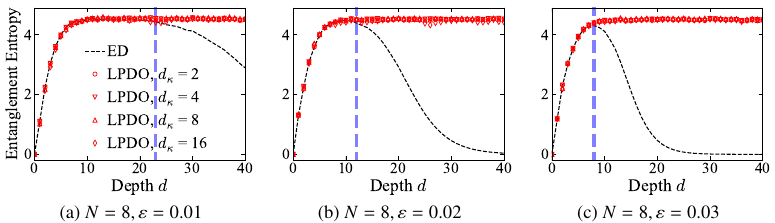}
    \caption{Entanglement entropy dynamics of noisy quantum circuits with $N=8$ and different $\varepsilon$ for ED and LPDO with $\chi=16$ and different $d_{\kappa}$.
    The blue dashed lines indicate the positions where $\rm EE_{ED}$ and $\rm EE_{LPDO}$ ($\chi=16$, $d_{\kappa}=2$) deviate by more than $3\%$.}
    \label{Fig: Dk}
\end{figure}

In the main text, we have compared the LPDO simulation results for different $\chi$ with a fixed $d_{\kappa}=2$.
Here, we further investigate the influence of $d_{\kappa}$ on the performance of the LPDO structure while maintaining the virtual dimension fixed at $\chi=16$ in Fig.~\ref{Fig: Dk}.
Remarkably, the results demonstrate a consistent trend across different values of $d_{\kappa}$, suggesting that the failure of LPDO near the quantum-classical crossover point cannot be alleviated only by increasing the Kraus dimension $d_{\kappa}$.
It is notable that if one takes a bond dimension of $D=\chi^2=256$, MPO simulation can provide exact results for the entire density matrix, implying a fundamental gap between LPDO with $\chi$ (even with a sufficiently large $d_{\kappa}$) and MPO with the corresponding $D=\chi^2$.

\begin{figure}[H]
    \centering
    \includegraphics[width=0.72\linewidth]{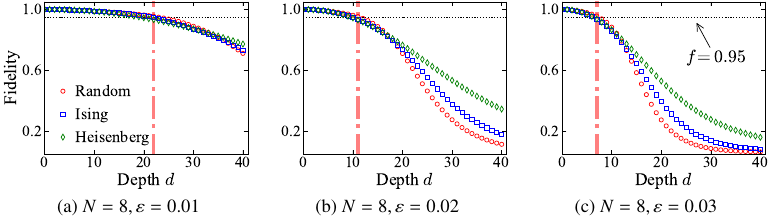}
    \caption{Fidelity between ED and LPDO with $d_{\kappa}=2$ and $\chi=16$, for noisy quantum circuits with $N=8$ and different $\varepsilon$, including random circuits, time evolution of Ising model, and that of Heisenberg model, with a time step of $\delta t=0.1$.
    The black horizontal dashed line indicates the threshold of $f=0.95$.
    The red dashed lines indicate the positions where $f_{\rm LPDO}$ for random gates drops below $0.95$.}
    \label{Fig: Fid}
\end{figure}

In addition to random quantum circuits, here we conduct numerical simulations for the quantum circuit to realize two typical spin models including the critical Ising model and the antiferromagnetic Heisenberg model, with a time step of $\delta t=0.1$.
The Hamiltonian of the Ising model is
\begin{align}
    H = -\sum_{i}Z_iZ_{i+1}-g\sum_i X_i,
\end{align}
with the critical point at $g=1$.
The Hamiltonian of the Heisenberg model is
\begin{align}
    H = \sum_{i}\bm{S}_i\cdot \bm{S}_{i+1}=\sum_{i}S_i^xS_{i+1}^x+S_i^yS_{i+1}^y+S_i^zS_{i+1}^z.
\end{align}
We still use the circuit configuration depicted in Fig.~\ref{Fig: Circuit}, where each two-qubit gate is chosen as
\begin{align}
    U = \exp{\left[-i\delta t\left(-Z_i Z_{i+1}-\frac{1}{2}X_i-\frac{1}{2}X_{i+1}\right)\right]}
\end{align}
for Ising model and 
\begin{align}
    U = \exp{\left[-i\delta t\left(S_i^xS_{i+1}^x+S_i^yS_{i+1}^y+S_i^zS_{i+1}^z\right)\right]}
\end{align}
for Heisenberg model.
The fidelity between ED and LPDO simulation is compared in Fig.~\ref{Fig: Fid} for different unitary gates, demonstrating a similar overall trend and a close optimal depth determined by the threshold $f=0.95$.

So far our focus has been on `strong-noise' scenarios, where the error rates for two-qubit gates range from $\varepsilon=0.01$ to $0.03$.
In the near future, as quantum hardware improves, we expect LPDO to demonstrate superior performance in simulating noisy quantum circuits than MPO in the `weak-noise' regime, especially in scenarios involving the time evolution of specific quantum systems that may show more quantum advantage rather than completely random circuits.
As an illustrative example, we conduct numerical simulations for the time evolution of the critical Ising model with a low error rate.

\begin{figure}[H]
    \centering
    \includegraphics[width=0.55\linewidth]{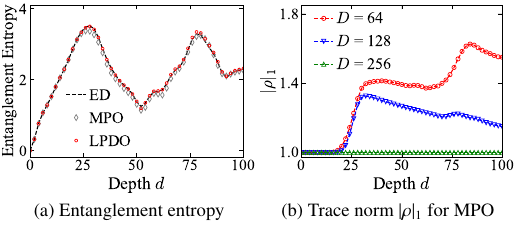}
    \caption{Time evolution for critical Ising model with $N=8$, $\delta t=0.1$, and $\varepsilon=0.001$.
    (a) Dynamics of entanglement entropy for ED, MPO with $D=64$, and LPDO with $\chi=16$ and $d_{\kappa}=2$.
    (b) Trace norm $|\rho|_1$ for MPO with different $D$.}
    \label{Fig: Ising}
\end{figure}

The time evolution of EE is illustrated in Fig.~\ref{Fig: Ising}(a), with a time step of $\delta t=0.1$ and a fixed error rate of $\varepsilon=0.001$, significantly lower than $\delta t$.
Our findings demonstrate that an LPDO with $\chi=16$ and $d_{\kappa}=2$ accurately captures the dynamics of entanglement.
In contrast, a truncated MPO with $D=64$ (note that $D=256$ is exact for $N=8$) introduces some errors.
Furthermore, we calculate the trace norm of the density matrix formed by the MPO, normalized as $\mathrm{Tr}(\rho)=1$, for various $D$ in Fig.~\ref{Fig: Ising}(b).
These results indicate that after truncation, the MPO no longer satisfies the condition of positivity and thus does not represent a physical density matrix.
Interestingly, the peaks of `negativity' observed in Fig.~\ref{Fig: Ising}(b) coincide with those of EE shown in Fig.~\ref{Fig: Ising}(a), revealing a deep relationship between these two phenomena.

We briefly discuss the physical implications of the observations above, focusing on the appropriate choice of ansatz in different scenarios.
In the case of strong noise, the LPDO simulation accurately captures dynamics for a limited number of layers, whereas the MPO simulation provides greater accuracy throughout the circuit at the expense of violating the positivity condition.
In contrast, LPDO demonstrates superior performance in scenarios with lower error rates, offering advantages in terms of both efficiency and accuracy.
Specifically, the number of parameters is $N\chi^2d_pd_{\kappa}$ for LPDO, compared to $ND^2d_p$ for MPO, which is approximately an order of magnitude smaller for the case studied here.
\end{document}